# Self-Dual Yang-Mills Multiplet
# in Three Dimensions Coupled to Supergravity

Roy MONTALVO, Hitoshi NISHINO,[1)] and Subhash RAJPOOT[2)]

*Department of Physics & Astronomy*
*California State University*
*1250 Bellflower Boulevard*
*Long Beach, CA 90840*

**Abstract**

We couple a recently-established $N = 1$ globally supersymmetric self-dual Yang-Mills multiplet in three dimensions to supergravity. This becomes possible due to our previous result on globally supersymmetric formulation based on a compensator multiplet. We further couple the self-dual vector to a supersymmetric $\sigma$-model on the coset $SO(8,n)/SO(8) \times SO(n)$ *via* minimal couplings for an arbitrary gauged subgroup $H_0 \subset SO(8) \times SO(n)$. A corresponding superspace formulation is also presented.

---

[1)] E-Mail: hnishino@csulb.edu
[2)] E-Mail: rajpoot@csulb.edu

## 1. Introduction

The concept of 'self-duality' for an Abelian vector in three dimensions (3D) was first introduced in [1], dictated by the relationship[3]

$$\tfrac{1}{2}\epsilon_\mu{}^{\rho\sigma} F_{\rho\sigma} \doteq m A_\mu \ . \tag{1.1}$$

As is usual for a vector field in any dimensions, the original physical degrees of freedom for $A_\mu$ is $3 - 2 = 1$, after the deduction of $2$ by the gauge fixing the longitudinal and time components. On the other hand, a repeated use of eq. (1.1) leads to

$$F_{\mu\nu} \doteq - m\epsilon_{\mu\nu}{}^\rho A_\rho \quad \Longrightarrow \quad \partial_\nu F^{\mu\nu} \doteq - m^2 A_\mu \ . \tag{1.2}$$

This also implies that $A$ is divergence-less

$$\partial_\mu A^\mu \doteq 0 \ , \tag{1.3}$$

and therefore (1.2) implies the *massive* vector field equation

$$(\partial_\nu^2 - m^2) A_\mu \doteq 0 \ . \tag{1.4}$$

Hence, the physical degrees of freedom should be $3 - 1 = 2$ as in the case for a massive vector instead of the massless one with $3 - 2 = 1$ degree of freedom. However, these two massive degrees of freedom are again halved due to the self-duality condition (1.1), leaving only one degree of freedom after all [1]. Similar treatments in general odd dimensions are also given in [1].

We have recently generalized the supersymmetric Abelian result in [1] to non-Abelian gauge groups, *i.e.*, we have presented a globally $N = 1$ supersymmetric self-dual Yang-Mills multiplet in 3D [2]. The key ingredient was to introduce the compensator scalar multiplet that makes the whole system gauge invariant, even though gauge symmetry is a 'fake' symmetry. We have also succeeded in the corresponding superspace formulation, and its coupling to supersymmetric Dirac-Born-Infeld action [2].

Since all of these results are based on global $N = 1$ supersymmetry, the next natural step is to generalize them to local supersymmetry. In the present paper, we accomplish the coupling of our $N = 1$ globally supersymmetric self-dual Yang-Mills [2] to $N = 1$ supergravity [3]. Thanks to the compensator multiplet, the supergravity coupling works in a straightforward manner as in a conventional supergravity theory [4], such as Noether couplings at the cubic order, and quartic couplings which shows the internal consistency of the system. We next couple the self-dual Yang-Mills multiplet to a $\sigma$-model for the coset $G/H \equiv SO(8,n)/SO(8) \times SO(n)$ *via* minimal coupling for an arbitrary subgroup $H_0 \subset H$. Subsequently, we reformulate the some of these couplings in terms of superspace language.

---

[3] We use the symbol $\doteq$ for a field equation distinguished from an algebraic one in our paper.

The motivations of our present work can be now summarized into two items:

(i) The coupling of $N = 1$ globally supersymmetric system to $N = 1$ supergravity is the next natural (and in a sense imperative) step.

(ii) By coupling to $N = 1$ supergravity with all the quartic terms, we will see the classical consistency of our system.

We stress that the item (ii) is for *classical* consistency. Because the quantum consistency of our model might be problematic, due to the compensator scalar involved. However, there are two main reasons for our optimism for quantum behavior of our model: (1) The consistent coupling to supergravity provides a good support also for quantum consistency. For example, type IIA massive supergravity in 10D [5] has a 1-form (vector) field playing a role of a compensator for a 2-form tensor field. Type IIA massive supergravity has a good quantum behavior based on superstring theory. Even though our model is not based on superstring, type IIA massive theory [5] is an encouraging example to deal with compensators. (2) Thanks to local supersymmetry inherent in the system, we expect that quantum behaviors will be improved compared with non-supersymmetric systems. In fact, we have seen such as suppressed quadratic divergences, as well as finite supersymmetric theories.

At any rate, the quantum-level consistency is outside the scope of the present work. As such, we do not address this question here.

## 2. Preliminaries on Compensators

We review the procedure of describing the compensators for an arbitrary non-Abelian gauge group $H_0$ [2].[4)] We first introduce the compensator scalar field in the adjoint representation $\varphi \equiv \varphi^I T^I$, where $T^I$ ($I = 1, 2, \cdots, \dim H_0$) are the anti-hermitian generators, satisfying the commutator

$$[T^I, T^J] = f^{IJK} T^K \ , \tag{2.1}$$

---

[4)] We call this gauge group $H_0$ instead of $G$ which will be used for $G \equiv SO(8,n)$ for a $\sigma$-model.

with the usual structure constant $f^{IJK}$. Relevantly, the main definitions and important relationships in our previous paper [2] are summarized as follows:

$$F_{\mu\nu} \equiv \partial_\mu A_\nu - \partial_\nu A_\mu + m[A_\mu, A_\nu] \ , \tag{2.2a}$$

$$D_\mu e^\varphi \equiv \partial_\mu e^\varphi + mA_\mu e^\varphi \ , \quad P_\mu \equiv (D_\mu e^\varphi)e^{-\varphi} \ , \tag{2.2b}$$

$$D_{[\mu} P_{\nu]} = +\tfrac{1}{2}mF_{\mu\nu} + \tfrac{1}{2}[P_\mu, P_\nu] \ . \tag{2.2c}$$

We sometimes omit adjoint indices, whenever there is no ambiguity involved. The gauge-coupling constant $m$ for $H_0$ has the dimension of mass in 3D, because we assign the mass dimension 0 (or 1/2) to the bosons (or fermions) [2]. The finite gauge transformation properties of these quantities for the local $H_0$ group have been well known [6][7]

$$(e^\varphi)' = e^{-\Lambda} e^\varphi \ , \quad (e^{-\varphi})' = e^{-\varphi} e^\Lambda \ , \quad (D_\mu e^\varphi)' = e^{-\Lambda}(D_\mu e^\varphi) \ , \tag{2.3a}$$

$$A_\mu{}' = m^{-1} e^{-\Lambda} \partial_\mu e^\Lambda + e^{-\Lambda} A_\mu e^\Lambda \ , \quad F_{\mu\nu}{}' = e^{-\Lambda} F_{\mu\nu} e^\Lambda \ , \tag{2.3b}$$

where $\Lambda \equiv \Lambda^I(x) T^I$ are $x$-dependent finite local gauge transformation parameters. All the terms in (2.3) are Lie-ring valued, as the adjoint indices are suppressed.

We can now depict the role of the compensator scalars through the 'toy' lagrangian[5]

$$\mathcal{L}_1 = -\tfrac{1}{4}(F_{\mu\nu}{}^I)^2 - \tfrac{1}{2}(P_\mu{}^I)^2 \ . \tag{2.4}$$

The new field redefined by

$$\widetilde{A}_\mu \equiv e^{-\varphi} A_\mu e^\varphi + m^{-1} e^{-\varphi}(\partial_\mu e^\varphi) = m^{-1} e^{-\varphi} P_\mu e^\varphi \ , \tag{2.5}$$

and its field strength do *not* transform [8][7]: $\widetilde{A}_\mu{}' = \widetilde{A}_\mu$, $\widetilde{F}_{\mu\nu}{}' = \widetilde{F}_{\mu\nu}$. The original lagrangian (2.4) can now be completely rewritten as the following lagrangian, where the exponential factor $e^{\pm\varphi}$ are entirely absent:

$$\mathcal{L}_1 = -\tfrac{1}{4}(\widetilde{F}_{\mu\nu}{}^I)^2 - \tfrac{1}{2}m^2(\widetilde{A}_\mu{}^I)^2 \ . \tag{2.6}$$

As usual in compensator formulations, the original kinetic term for $\varphi$ is now reduced to the mass term of $\widetilde{A}_\mu$ [8][7], and the original gauge invariance is no longer manifest.

Instead of the $F_{\mu\nu}^2$-term in (2.4), consider now the supersymmetric Chern-Simons lagrangian [9] with an additional mass parameter $\mu$:

$$\begin{aligned}\mathcal{L}_2 &\equiv \tfrac{1}{4}\mu\,\epsilon^{\mu\nu\rho}(F_{\mu\nu}{}^I A_\rho{}^I - \tfrac{1}{3} f^{IJK} A_\mu{}^I A_\nu{}^J A_\rho{}^K) - \tfrac{1}{2}(P_\mu{}^I)^2 \\ &= \tfrac{1}{4}\mu\,\epsilon^{\mu\nu\rho}(\widetilde{F}_{\mu\nu}{}^I \widetilde{A}_\rho{}^I - \tfrac{1}{3} f^{IJK} \widetilde{A}_\mu{}^I \widetilde{A}_\nu{}^J \widetilde{A}_\rho{}^K) - \tfrac{1}{2} m^2 (\widetilde{A}_\mu{}^I)^2 \ . \end{aligned} \tag{2.7}$$

---

[5] Our metric in this paper is $(\eta_{\mu\nu}) = $ diag. $(-,+,+)$.

This yields the $A$-field equation, or equivalently the $\widetilde{A}$-field equation

$$\tfrac{1}{2}\mu\,\epsilon^{\mu\nu\rho}F_{\nu\rho}{}^I \doteq mP_\mu{}^I \ , \qquad \tfrac{1}{2}\mu\,\epsilon^{\mu\nu\rho}\widetilde{F}_{\nu\rho}{}^I \doteq m^2\widetilde{A}_\mu{}^I \ . \tag{2.8}$$

The latter is nothing but the self-duality (1.1), if $\mu \equiv m$, and $\widetilde{A}_\mu$ is identified with $A_\mu$.

As we have seen here, the advantage of the compensator formulation is to use the gauge invariance to fix lagrangians easily because only limited lagrangian terms are allowed under the gauge invariance of the action.

## 3. Locally Supersymmetric Self-Dual Yang-Mills Multiplet

We are now ready to couple the $N=1$ globally supersymmetric self-dual Yang-Mills to supergravity. Our field content is the multiplet of $N=1$ supergravity $(e_\mu{}^m, \psi_\mu)$, the Yang-Mills multiplet $(A_\mu{}^I, \lambda^I)$ and the compensator scalar multiplet $(\varphi^I, \chi^I)$.

Even though the self-dual Yang-Mills multiplet in 3D implies the absence of the usual kinetic terms starting with $-(1/4)(F_{\mu\nu}{}^I)^2$, we first consider such kinetic terms, as the general option. The coupling procedure then is similar to the routine Noether couplings, together with the conventional method for quartic fermion terms [4]. Thanks to the compensator multiplet, the coupling procedure is simplified.

Our first result is summarized by the total action $I_3 \equiv I_{\text{SG}} + I_{\text{VM}} + I_{\text{SM}} + I_{\text{CS}} + I_{\text{CC}}$, where $I_{\text{SG}}$ is the kinetic terms for supergravity, $I_{\text{VM}}$ is for the kinetic terms for the Yang-Mills multiplet, $I_{\text{SM}}$ is for the kinetic terms for the compensator scalar multiplet, $I_{\text{CS}}$ is the supersymmetric Chern-Simons terms, and $I_{\text{CC}}$ is for a cosmological constant. Their corresponding lagrangians are respectively,

$$e^{-1}\mathcal{L}_{\text{SG}} = -\tfrac{1}{4}R - e^{-1}\epsilon^{\mu\nu\rho}(\overline{\psi}_\mu D_\nu(\omega)\psi_\rho) \ , \tag{3.1a}$$

$$e^{-1}\mathcal{L}_{\text{VM}} = -\tfrac{1}{4}(F_{\mu\nu}{}^I)^2 - \tfrac{1}{2}(\overline{\lambda}^I \slashed{D}(\omega,A)\lambda^I) - \tfrac{1}{4}(\overline{\psi}_\mu \gamma^{\rho\sigma}\gamma^\mu \lambda^I)(F_{\rho\sigma}{}^I + \widehat{F}_{\rho\sigma}{}^I)$$
$$\qquad - m(\overline{\lambda}^I \chi^I) + \tfrac{1}{8}(\overline{\lambda}^I \lambda^I)^2 \ , \tag{3.1b}$$

$$e^{-1}\mathcal{L}_{\text{SM}} = -\tfrac{1}{2}(P_\mu{}^I)^2 - \tfrac{1}{2}(\overline{\chi}^I \slashed{D}\chi^I) + \tfrac{1}{48}h^{IJ,KL}(\overline{\chi}^I\chi^K)(\overline{\chi}^J\chi^L)$$
$$\qquad + \tfrac{1}{2}(\overline{\psi}_\mu \gamma^\nu \gamma^\mu \chi^I)(P_\nu{}^I + \widehat{P}_\nu{}^I) - \tfrac{1}{8}(\overline{\chi}^I\chi^I)^2 + \tfrac{1}{4}(\overline{\lambda}^I\lambda^I)(\overline{\chi}^J\chi^J) \ , \tag{3.1c}$$

$$e^{-1}\mathcal{L}_{\text{CS}} = +\tfrac{1}{4}\mu\,e^{-1}\epsilon^{\mu\nu\rho}(F_{\mu\nu}{}^I A_\rho{}^I - \tfrac{1}{3}mf^{IJK}A_\mu{}^I A_\nu{}^J A_\rho{}^K) - \tfrac{1}{2}\mu\,(\overline{\lambda}^I\lambda^I) \ , \tag{3.1d}$$

$$e^{-1}\mathcal{L}_{\text{CC}} = +M(\overline{\psi}_\mu \gamma^{\mu\nu}\psi_\nu) + 2M^2 + \tfrac{1}{2}M(\overline{\lambda}^I\lambda^I) + \tfrac{1}{2}M(\overline{\chi}^I\chi^I) \ . \tag{3.1e}$$

The constant $h$'s in $\mathcal{L}_{\mathrm{SM}}$ is defined in terms of the structure constant $f^{IJK}$ of $H_0$:

$$h^{IJ,KL} \equiv f^{IJM} f^{MKL} \ . \tag{3.2}$$

The covariant derivative $\mathcal{D}$ acts on $\chi$ as in the globally supersymmetric case [2] except for the Lorentz connection term:

$$\mathcal{D}_\mu \chi^I \equiv -\tfrac{1}{2} m f^{IJK} P_\mu{}^J \chi^K + \partial_\mu \chi^I - \tfrac{1}{4} \omega_\mu{}^{rs} \gamma_{rs} \chi^I \ . \tag{3.3}$$

The field strength $F_{\mu\nu}{}^I$ is the same as (2.2a), while all the *hatted* field strengths are their supercovariantizations [4], defined by

$$\widehat{F}_{\mu\nu}{}^I \equiv F_{\mu\nu}{}^I - 2(\overline{\psi}_{[\mu} \gamma_{\nu]} \lambda^I) \ , \tag{3.4a}$$

$$\widehat{P}_\mu{}^I \equiv [(\widehat{D}_\mu e^\varphi) e^{-\varphi}]^I \equiv [\{\partial_\mu e^\varphi - (\overline{\psi}_\mu \chi) e^\varphi + m A_\mu e^\varphi\} e^{-\varphi}]^I \ . \tag{3.4b}$$

Our total action $I_3$ is invariant under local $N=1$ supersymmetry

$$\delta_Q e_\mu{}^m = +2(\overline{\epsilon} \gamma^m \psi_\mu) \ , \tag{3.5a}$$

$$\delta_Q \psi_\mu = +D_\mu(\widehat{\omega}) \epsilon + M(\gamma_\mu \epsilon) \ , \tag{3.5b}$$

$$\delta_Q A_\mu{}^I = +(\overline{\epsilon} \gamma_\mu \lambda^I) \ , \tag{3.5c}$$

$$\delta_Q \lambda^I = -\tfrac{1}{2} (\gamma^{\mu\nu} \epsilon) \widehat{F}_{\mu\nu}{}^I \ , \tag{3.5d}$$

$$\delta_Q e^\varphi = +(\overline{\epsilon} \chi^I) e^\varphi \ , \tag{3.5e}$$

$$\delta_Q \chi^I = +(\gamma^\mu \epsilon) \left[ \widehat{P}_\mu{}^I - \tfrac{1}{4} f^{IJK} (\overline{\chi}^J \gamma_\mu \chi^K) \right] \ . \tag{3.5f}$$

As usual, $\widehat{\omega}_\mu{}^{rs} \equiv \widehat{\omega}_\mu{}^{rs}(e,\psi)$ is the Lorentz connection with the $\psi$-torsion included [4].

Some remarks are in order. First, the normalization of the coefficient for the gravitino kinetic term is the unit strength instead of $1/2$, due to the commutation relations $[\delta_Q(\epsilon_1), \delta_Q(\epsilon_2)] = +2(\overline{\epsilon}_2 \gamma^m \epsilon_1) P_m$, also reflected in the coefficient '+2' in (3.5a).

Second, for a *self-dual* VM, the kinetic lagrangian $\mathcal{L}_{\mathrm{VM}}$ should be dropped, and the total action should be $I_4 \equiv I_{\mathrm{SDVM}} \equiv I_{\mathrm{SG}} + I_{\mathrm{SM}} + I_{\mathrm{CS}} + I_{\mathrm{CC}}$. Accordingly, when $\mathcal{L}_{\mathrm{VM}}$ is dropped, the self-duality condition in the globally supersymmetric case is now generalized to locally supersymmetric equation

$$\tfrac{1}{2} \mu e^{-1} \epsilon_\mu{}^{\rho\sigma} \widehat{F}_{\rho\sigma} \doteq m \widehat{P}_\mu - \tfrac{1}{4} m f^{IJK} (\overline{\chi}^J \gamma_\mu \chi^K) \ . \tag{3.6}$$

This is nothing but the locally supersymmetric generalization of the gauge-covariantized form (2.8) of the self-duality (1.1). Due to the free parameter $\mu$, we have more freedom than the special case $\mu = m$. The globally supersymmetric version in [2] can be also re-obtained by deleting the gravitino and graviton fields.

Third, the normalization of the terms in $\mathcal{L}_{\text{CS}}$ has been chosen, such that the self-duality condition (3.6) easily recovers the non-supersymmetric case (1.1). However, for any gauge group whose $\pi_3$-mapping is non-trivial, such as [10]

$$\pi_3(H_0) = \begin{cases} \mathbb{Z} & (\text{for } H_0 = A_i, \ B_i, \ C_i, \ D_i \ (i \geq 2, \ H_0 \neq D_2), \ F_4, \ G_2, \ E_6, \ E_7, \ E_8) \ , \\ \mathbb{Z} \oplus \mathbb{Z} & (\text{for } H_0 = SO(4)) \ , \\ 0 & (\text{for } H_0 = U(1)) \ , \end{cases} \tag{3.7}$$

the constant $\mu$ should be quantized as

$$\mu = \frac{nm^2}{\pi} \quad (n = 0, \ \pm 1, \ \pm 2, \ \cdots) \ . \tag{3.8}$$

Fourth, the supersymmetric cosmological constant term $\mathcal{L}_{\text{CC}}$ can be obtained by the routine procedure starting with the cosmological constant proportional to $M^2$ and the gravitino mass term proportional to $M(\overline{\psi}_\mu \gamma^{\mu\nu} \psi_\nu)$. The positive definite signature $M^2 > 0$ implies the anti-de Sitter space-time in 3D. The new feature here is that this cosmological constant induces the mass terms both for the gaugino $\lambda$ and the fermionic partner $\chi$ in the compensator multiplet. The mass terms of spin 1/2 fields $\lambda$ and $\chi$ induced by the cosmological constant is not peculiar to this system, but it is rather universal in other dimensions, *e.g.*, type IIA supergravity [11].

Fifth, each lagrangian in (3.1) is *not* by itself invariant. For example, $I_{\text{SDVM}}$ defined above is invariant, but *not* each lagrangian in $I_{\text{SDVM}}$. We also need a special care, when dropping some lagrangians (3.1a) through (3.1e) in $I_3$, in order to maintain the invariance of the resulting total action. For example, when we drop $\mathcal{L}_{\text{CC}}$, we have to drop the $M$-term in the transformation (3.5b), setting $M = 0$ everywhere in the system.

Sixth, there are three mass terms for the $\lambda$ and $\chi$-fields:

$$\tfrac{1}{2}(\overline{\lambda}^I, \overline{\chi}^I) \begin{pmatrix} M - \mu & -m \\ -m & M \end{pmatrix} \begin{pmatrix} \lambda^I \\ \chi^I \end{pmatrix} \ , \tag{3.9}$$

whose eigenvalues $\mathcal{M}$ are computed to be

$$\begin{aligned} \mathcal{M} &= M - \tfrac{\mu}{2} \pm \sqrt{\tfrac{\mu^2}{4} + m^2} \\ &= M - \frac{nm^2}{2\pi} \pm \sqrt{\frac{n^2 m^4}{4\pi^2} + m^2} \quad (n = 0, \ \pm 1, \ \pm 2, \ \cdots) \ , \end{aligned} \tag{3.10}$$

due to (3.8) for a non-Abelian group in the unit of $\kappa = 1$. If we further impose an additional condition between these eigenvalues, such as one of them to be zero, then the cosmological constant $\Lambda \equiv 2M^2$ itself and/or the gauge coupling constant $m$ will be quantized.

## 4. Coupling to $\sigma$-Model on $SO(8,n)/SO(8) \times SO(n)$

As another example of non-trivial couplings of our self-dual Yang-Mills multiplet in 3D, we introduce the $SO(8,n)/SO(8) \times SO(n)$ $\sigma$-model originally developed by [12] for extended $N = 8$ supergravity, and applied also to $N = 1$ supergravity in 3D in our previous paper [13].

We choose the coset $G/H \equiv SO(8,n)/SO(8) \times SO(n)$ because of its non-trivial and rich structure. In particular, since $n = 1, 2, \cdots$ can be general, the size of this coset can be arbitrarily large with many potential applications. Moreover, the existence of two groups in $H = SO(8) \times SO(n)$ makes system non-trivial, serving as a template for more complicated cosets. Despite such a rich coset structure, the number of supersymmetry remains to be $N = 1$, as will be elucidated in the supersymmetry transformation rule (4.8) and also in [13]. This is in contrast with, e.g., $N = 2$ hyper Kähler manifold in 4D [14], where extended $N \geq 2$ supersymmetries are required.

The new multiplet introduced is the $\sigma$-model multiplet $(\phi^{\underline{\alpha}}, \rho^{Aa})$ in addition to the Yang-Mills multiplet $(A_\mu{}^I, \lambda^I)$, the compensator multiplet $(\varphi^I, \chi^I)$, and that of supergravity $(e_\mu{}^m, \psi_\mu)$. The latter three are the same multiplets introduced in the previous section. The scalars $\phi^{\underline{\alpha}}$ are the coordinates of the coset $G/H \equiv SO(8,n)/SO(8) \times SO(n)$ [13]. The vector $A_\mu{}^I$ is supposed to satisfy the self-duality condition (1.1) with its supersymmetric generalizations. The indices $\underline{\alpha}, \underline{\beta}, \cdots = 1, 2, \cdots, 8n = \dim(G/H)$ are for the curved indices of the manifold $G/H$, while $A, B, \cdots = 1, 2, \cdots, 8$ are for the vectorial $\mathbf{8}_V$ of $SO(8)$,[6] while $a, b, \cdots = 1, 2, \cdots, n$ are for the $\mathbf{n}$ of $SO(n)$. The indices $I, J, \cdots = 1, 2, \cdots, \dim H_0$ are for the adjoint representation for $H_0$ which is an arbitrary gauged subgroup of $H \equiv SO(8) \times SO(n)$.

As usual for supersymmetric $\sigma$-models, we introduce the vielbein $V_{\underline{\alpha}}{}^{Aa}$ and its inverse $V_{Aa}{}^{\underline{\alpha}}$ on $G/H$ [12][13]. For the gauging of $H_0 \subset H$, we introduce the Killing vectors $\xi^{\underline{\alpha} I}$ into the covariant derivative of the coordinates $\phi^{\underline{\alpha}}$ [14][15]:

$$\mathcal{D}_\mu \phi^{\underline{\alpha}} \equiv \partial_\mu \phi^{\underline{\alpha}} - m A_\mu{}^I \xi^{\underline{\alpha} I} \ . \tag{4.1}$$

---

[6] In this paper, we assign the vectorial $\mathbf{8}_V$ representation for these indices instead of the spinorial $\mathbf{8}_S$ in [13], in order to simplify supergravity couplings.

Here we have to use $m$ for the gauge coupling constant for the total consistency. Eq. (4.1) is equivalent to the expression in terms of the coset representative $\mathcal{V}$ [6][14][12][16][13]:

$$\mathcal{V}^{-1}\mathcal{D}_\mu \mathcal{V} \equiv \mathcal{V}^{-1}\partial_\mu \mathcal{V} + mA_\mu{}^I \mathcal{V}^{-1}T^I \mathcal{V}$$
$$= (\mathcal{D}_\mu \phi^{\underline{\alpha}})\left(V_{\underline{\alpha}}{}^{Aa}Y^{Aa} + \tfrac{1}{2}Q_{\underline{\alpha}}{}^{AB}X^{AB} + \tfrac{1}{2}Q_{\underline{\alpha}}{}^{ab}X^{ab}\right) \ . \tag{4.2}$$

In the contractions among the indices $A, B, \cdots$ or $a, b, \cdots$, we always use superscripts, because the corresponding metrics are all positive definite. The $T^I$'s are the generators of the arbitrary gauged group $H_0 \subset H \equiv SO(8) \times SO(n)$, and $Y^{Aa}$'s are the generators on the coset $G/H$, while $X^{AB}$ (or $X^{ab}$) are the generators of $SO(8)$ (or $SO(n)$), satisfying their algebras

$$\begin{aligned}
[X^{AB}, X^{CD}] &= +\delta^{BC}X^{AD} - \delta^{AC}X^{BD} - \delta^{BD}X^{AC} + \delta^{AD}X^{BC} \ , \\
[X^{ab}, X^{cd}] &= +\delta^{bc}X^{ad} - \delta^{ac}X^{bd} - \delta^{bd}X^{ac} + \delta^{ad}X^{bc} \ , \\
[X^{AB}, Y^{Dd}] &= +\delta^{BD}Y^{Ad} - \delta^{AD}Y^{Bd} \ , \\
[X^{ab}, Y^{Dd}] &= +\delta^{bd}Y^{Da} - \delta^{ad}Y^{Db} \ , \\
[Y^{Ab}, Y^{Cd}] &= +\delta^{AC}X^{bd} + \delta^{bd}X^{AC} \ .
\end{aligned} \tag{4.3}$$

Accordingly, the Killing vectors satisfy the relationships

$$D^{Aa}(Q)\xi^{BbI} \equiv V^{Aa\underline{\alpha}}(\partial_{\underline{\alpha}}\xi^{BbI} + Q_{\underline{\alpha}}{}^{BC}\xi^{CbI} + Q_{\underline{\alpha}}{}^{bc}\xi^{BcI}) = \delta^{ab}C^{ABI} + \delta^{AB}C^{abI} \ , \tag{4.4}$$

where $\xi^{BbI} \equiv V_\beta{}^{Bb}\xi^{\beta I}$, and the $C$'s defined by

$$C^{ABI} \equiv Q_{\underline{\alpha}}{}^{AB}\xi^{\underline{\alpha}I} \ , \qquad C^{abI} \equiv Q_{\underline{\alpha}}{}^{ab}\xi^{\underline{\alpha}I} \ , \tag{4.5}$$

have been known [15] to be covariant both under the composite $\sigma$-model gauge transformations on $G/H$ and the gauged subgroup $H_0 \subset H$.

Our total action is now $I_5 \equiv \int d^4x \, \mathcal{L}_5$, where

$$\begin{aligned}
e^{-1}\mathcal{L}_5 =& -\tfrac{1}{4}R - e^{-1}\epsilon^{\mu\nu\rho}(\overline{\psi}_\mu D_\nu \psi_\rho) - \tfrac{1}{2}g_{\underline{\alpha}\underline{\beta}}g^{\mu\nu}(\mathcal{D}_\mu\phi^{\underline{\alpha}})(\mathcal{D}_\nu\phi^{\underline{\beta}}) - \tfrac{1}{2}(P_\mu{}^I)^2 \\
& -\tfrac{1}{2}(\overline{\chi}^I \slashed{D}\chi^I) - \tfrac{1}{2}(\overline{\rho}^{Aa}\gamma^\mu \slashed{D}\rho^{Aa}) - m(\overline{\lambda}^I\chi^I) + mV_{\underline{\alpha}}{}^{Aa}(\overline{\rho}^{Aa}\lambda^I)\xi^{\underline{\alpha}I} \\
& +\tfrac{1}{2}V_{\underline{\alpha}}{}^{Aa}(\overline{\psi}_\mu\gamma^\nu\gamma^\mu\rho^{Aa})(\mathcal{D}_\nu\phi^{\underline{\alpha}} + \widehat{\mathcal{D}}_\nu\phi^{\underline{\alpha}}) + \tfrac{1}{2}(\overline{\psi}_\mu\gamma^\nu\gamma^\mu\chi^I)(P_\nu{}^I + \widehat{P}_\nu{}^I) \\
& +\tfrac{1}{4}\mu\, e^{-1}\epsilon^{\mu\nu\rho}\left(F_{\mu\nu}{}^I A_\rho{}^I - \tfrac{1}{3}mf^{IJK}A_\mu{}^I A_\nu{}^J A_\rho{}^K\right) - \tfrac{1}{2}\mu(\overline{\lambda}^I\lambda^I) \\
& -\tfrac{1}{8}(\overline{\rho}^{Aa}\rho^{Aa})^2 - \tfrac{1}{6}(\overline{\rho}^{Aa}\gamma_\mu\rho^{Ba})^2 - \tfrac{1}{6}(\overline{\rho}^{Aa}\gamma_\mu\rho^{Ab})^2 + \tfrac{1}{4}(\overline{\rho}^{Aa}\rho^{Aa})(\overline{\lambda}^I\lambda^I) \\
& +\tfrac{1}{48}h^{IJ,KL}(\overline{\chi}^I\chi^K)(\overline{\chi}^J\chi^L) + \tfrac{1}{4}(\overline{\lambda}^I\lambda^I)(\overline{\chi}^J\chi^J) - \tfrac{1}{8}(\overline{\chi}^I\chi^I)^2 \ .
\end{aligned} \tag{4.6}$$

The $h$'s is defined formally by (3.2), but now the structure constant $f^{IJK}$ is for the gauged group $H_0 \in H \equiv SO(8) \times SO(n)$. As usual in supergravity in 3D [12][16][13], we adopt the 1.5-order formalism for the Lorentz connection $\omega_\mu{}^{rs}$ [4], such as $D_\mu \psi_\rho \equiv \partial_\mu \psi_\rho - (1/4)\omega_\mu{}^{rs}\gamma_{rs}\psi_\rho$, where the Lorentz connection $\omega$ is treated as an independent field, satisfying its algebraic field equation. The covariant derivative $\mathcal{D}_\mu$ on $\rho$ is defined by [13]

$$\mathcal{D}_\mu \rho^{Aa} \equiv +\partial_\mu \rho^{Aa} - \tfrac{1}{4}\omega_\mu{}^{rs}\gamma_{rs}\rho^{Aa} + Q_\mu{}^{AB}\rho^{Ba} + Q_\mu{}^{ab}\rho^{Ab} + mA_\mu{}^I (T^I)^{\underline{AB}}\rho^{\underline{B}} \ . \tag{4.7}$$

where $Q_\mu{}^{IJ} \equiv (\mathcal{D}_\mu \phi^{\underline{\alpha}})Q_{\underline{\alpha}}{}^{IJ}$ and $Q_\mu{}^{bc} \equiv (\mathcal{D}_\mu \phi^{\underline{\alpha}})Q_{\underline{\alpha}}{}^{bc}$ are the composite connections with their pullbacks, as usual [12][16][13]. The underlined indices $\underline{A}$, $\underline{B}$, $\cdots$ are for the pair of indices $A'a'$, $B'b'$, $\cdots$, e.g., $\rho^{\underline{B}} \equiv \rho^{B'b'}$, etc., where these *primed* indices $A'$, $B'$, $\cdots$ and $a'$, $b'$, $\cdots$ are the subgroups of the original indices $A$, $B$, $\cdots$ and $a$, $b$, $\cdots$, depending on the gauged subgroup $H_0$ on which $T^I$ acts non-trivially.

Our total action $I_5$ is invariant under local $N=1$ supersymmetry:

$$\delta_Q e_\mu{}^m = +2(\overline{\epsilon}\gamma^m \psi_\mu) \ , \tag{4.8a}$$

$$\delta_Q \psi_\mu = +D_\mu(\widehat{\omega})\epsilon \ , \tag{4.8b}$$

$$\delta_Q A_\mu{}^I = +(\overline{\epsilon}\gamma_\mu \lambda^I) \ , \tag{4.8c}$$

$$\delta_Q \lambda^I = -\tfrac{1}{2}(\gamma^{\mu\nu}\epsilon)\widehat{F}_{\mu\nu}{}^I \ , \tag{4.8d}$$

$$\delta_Q e^\varphi = +(\overline{\epsilon}\chi)e^\varphi \ , \tag{4.8e}$$

$$\delta_Q \chi^I = +(\gamma^\mu \epsilon)\left[\widehat{P}_\mu{}^I - \tfrac{1}{4}f^{IJK}(\overline{\chi}^J \gamma_\mu \chi^K)\right] \ , \tag{4.8f}$$

$$\delta_Q \phi^{\underline{\alpha}} = +V^{Aa\underline{\alpha}}(\overline{\epsilon}\rho^{Aa}) \ , \tag{4.8g}$$

$$\delta_Q \rho^{Aa} = +(\gamma^\mu \epsilon)V_{\underline{\alpha}}{}^{Aa}\widehat{\mathcal{D}}_\mu \phi^{\underline{\alpha}} - (\delta_Q \phi^{\underline{\alpha}})(Q_{\underline{\alpha}}{}^{AB}\rho^{Ba} + Q_{\underline{\alpha}}{}^{ab}\rho^{Ab}) \ . \tag{4.8h}$$

Note that we have $N=1$ supersymmetry, despite the coset $SO(8,n)/SO(8) \times SO(n)$, as the index structures in (4.8g) and (4.8h) show. As before, all the *hatted* field strengths are supercovariantized [4], e.g.,

$$\widehat{\mathcal{D}}_\mu \phi^{\underline{\alpha}} \equiv \mathcal{D}_\mu \phi^{\underline{\alpha}} - V^{Aa\underline{\alpha}}(\overline{\psi}_\mu \rho^{Aa}) \ . \tag{4.9}$$

Some remarks are in order. First, since we are dealing with the self-dual Yang-Mills multiplet, we do not need the kinetic terms for this multiplet. Accordingly, terms in $\mathcal{L}_{\text{VM}}$, such as the Noether term $\overline{\psi}\gamma\gamma\lambda F$, or the quartic terms such as $\psi^2 \lambda^2$ or $\lambda^4$ are all absent.

Second, as far as the $\sigma$-model part is concerned, these couplings are essentially the same as our previous paper [13]. The only differences are with the coefficients, caused by the notational change from [13], such as the metric signature, the $\mathbf{8}_V$ of $SO(8)$, or the scaling of supersymmetry commutator algebra.

Third, we see that the *self-dual* Yang-Mills vector can be coupled to the $\sigma$-model on the coset $SO(8,n)/SO(8) \times SO(n)$ consistently with supersymmetry. In particular, the $A_\mu$-field equation yields the $\sigma$-model corrected version of the self-duality condition (3.6), as

$$\tfrac{1}{2}\mu\epsilon_\mu{}^{\rho\sigma}\widehat{F}_{\rho\sigma}{}^I \doteq + m\widehat{P}_\mu{}^I - \tfrac{1}{4}mf^{IJK}(\overline{\chi}^J\gamma_\mu\chi^K) - m\xi_{\underline{\alpha}}{}^I\widehat{\mathcal{D}}_\mu\phi^{\underline{\alpha}} + \tfrac{1}{2}m(T^I)^{\underline{AB}}(\overline{\rho}^{\underline{A}}\gamma_\mu\rho^{\underline{B}})$$
$$- \tfrac{1}{2}mC^{ABI}(\overline{\rho}^{Aa}\gamma_\mu\rho^{Ba}) - \tfrac{1}{2}mC^{abI}(\overline{\rho}^{Aa}\gamma_\mu\rho^{Ab}) \ . \tag{4.10}$$

This result is also due to our previous formulation based on the compensator multiplet that simplifies supergravity couplings. Self-dual vectors coupled to $\sigma$-model have been presented also in the context of $N = 16$ gauged supergravity [16]. However, our coupling between a self-dual gauge field and a $\sigma$-model in 3D is the simplest one that can be used as a template for more applications related to $\sigma$-models.

Fourth, the invariance of the action $I_5$ can be confirmed as in usual supergravity. Aside from quartic terms, all the $m$-independent cubic terms are the usual routine computations. As for the $m$-dependent cubic terms, only $\sigma$-model dependent terms are the new contributions, categorized into four sectors (i) $m\rho F$, (ii) $m\lambda\mathcal{D}\phi$, (iii) $m\rho^2\lambda$, and (iv) $m\psi\lambda\rho$. To all of these sectors, the explicit term $m\overline{\rho}\lambda\xi$ in the lagrangian contributes, analogously to the usual gaugino-quark-squark mixing term [17]. Sector (i) comes from the variation of the gravitino in the Noether term $\overline{\psi}\rho\mathcal{D}\phi$ and the $m\overline{\rho}\lambda\xi$-term in the lagrangian. The former generates the commutator $[\mathcal{D}_\mu, \mathcal{D}_\nu]$ cancelling the like term from the variation of the latter. Sector (ii) comes from the minimal coupling of the $\phi$-kinetic term and the $m\overline{\rho}\lambda\xi$-term. Sector (iii) comes from the composite connections in the $\rho$-kinetic term, and the $m\overline{\rho}\lambda\xi$-term. Finally, sector (iv) comes from the $m\overline{\rho}\lambda\xi$-term and the Noether term $\overline{\psi}\rho\mathcal{D}\phi$. In these computations, the following $\sigma$-model related formulae are needed [6][14][15]:

$$\mathcal{D}_\mu\mathcal{D}_\nu\phi^{\underline{\alpha}} \equiv \partial_\mu\mathcal{D}_\nu\phi^{\underline{\alpha}} - mA_\mu{}^I(\partial_{\underline{\beta}}\xi^{\underline{\alpha}I})\mathcal{D}_\nu\phi^{\underline{\beta}} \ , \tag{4.11a}$$

$$\mathcal{L}_{\xi^J}\xi^{\beta I} = \xi^{\underline{\alpha}J}\partial_{\underline{\alpha}}\xi^{\underline{\beta}I} - \xi^{\underline{\alpha}I}\partial_{\underline{\alpha}}\xi^{\underline{\beta}J} + mf^{IJK}\xi^{\underline{\beta}K} = 0 \ , \tag{4.11b}$$

$$\mathcal{L}_{\xi^I}V_{\underline{\alpha}}{}^{Aa} = \xi^{\underline{\beta}I}\partial_{\underline{\beta}}V_{\underline{\alpha}}{}^{Aa} + (\partial_{\underline{\alpha}}\xi^{\underline{\beta}I})V_{\underline{\beta}}{}^{Aa} + m(T^I)^{\underline{AB}}V_{\underline{\alpha}}{}^{\underline{B}} = 0 \ , \tag{4.11c}$$

$$[\mathcal{D}_\mu, \mathcal{D}_\nu]\phi^{\underline{\alpha}} = -mF_{\mu\nu}{}^I\xi^{\underline{\alpha}I} \ , \tag{4.11d}$$

where $\mathcal{L}_{\xi^I}$ stands for a Lie derivative in the $\xi^{\underline{\alpha}I}$-direction.

Fifth, the remaining quartic terms in the lagrangian are also parallel to the previous section, or to the usual $N = 1$ supergravity [13] which do not need additional clarifications.

## 5. Superspace Reformulation

Once we have established component formulation of coupling between self-dual Yang-Mills based on compensator scalar multiplet, the next natural step is to reformulate in superspace [18]. Even though we do not include the $\sigma$-model multiplet and supergravity in this section, we already see highly non-trivial relationships needed for the mutual consistency of superfield equations.

Although a lagrangian formulation in superspace has been presented in our previous paper [2], we present here another superspace formulation based on Bianchi identities (BIds), which provides important formulae, as will be seen. The relevant superfield strengths are the Yang-Mills superfield strength $F_{AB}{}^I$, supertorsion $T_{AB}{}^C$ and supercurvature $R_{ABc}{}^d$ [18], together with our new superfield strength $P_A{}^I$:[7]

$$\tfrac{1}{2}\nabla_{[A}T_{BC)}{}^D - \tfrac{1}{2}T_{[AB|}{}^E T_{E|C)}{}^D - \tfrac{1}{4}R_{[AB|e}{}^f(\mathcal{M}_f{}^e)_{|C)}{}^D \equiv 0 \ , \tag{5.1a}$$

$$\tfrac{1}{2}\nabla_{[A}F_{BC)}{}^I - \tfrac{1}{2}T_{[AB|}{}^D F_{D|C)}{}^I \equiv 0 \ , \tag{5.1b}$$

$$\nabla_{[A}P_{B)}{}^I - T_{AB}{}^C P_C{}^I - f^{IJK} P_A{}^J P_B{}^K - m F_{AB}{}^I \equiv 0 \ . \tag{5.1c}$$

Here, the superfield strength $P_A{}^I$ is defined by [2]

$$P_A \equiv (\nabla_A e^\varphi) e^{-\varphi} \equiv (D_A e^\varphi + m A_A\, e^\varphi)\, e^{-\varphi} \ , \tag{5.2}$$

so that we have

$$P_\alpha{}^I = [\,(\nabla_\alpha e^\varphi) e^{-\varphi}\,]^I = -\chi_\alpha{}^I \ . \tag{5.3}$$

The superspace constraints satisfying the BIds (5.1) are given by

$$T_{\alpha\beta}{}^c = -2(\gamma^c)_{\alpha\beta} \ , \quad T_{\alpha\beta}{}^\gamma = T_{\alpha b}{}^c = T_{\alpha b}{}^\gamma = T_{ab}{}^c = 0 \ , \tag{5.4a}$$

---

[7] We use the indices $A \equiv (a,\alpha)$, $B \equiv (b,\beta)$, $\cdots$ in superspace, where $a, b, \cdots = 0, 1, 2$ (or $\alpha, \beta, \cdots = 1, 2$) are used for bosonic (or fermionic) superspace coordinates. Even though we use the same indices $A, B, \cdots$ both for superspace local coordinates and the $\mathbf{8}_V$ of $SO(8)$, or $a, b, \cdots$ both for bosonic superspace local coordinates and the $\mathbf{n}$ of $SO(n)$, they can be easily distinguished by the context. The antisymmetrization symbols are defined, e.g., by $M_{[AB)} \equiv M_{AB} - (-)^{AB} M_{BA}$. In superspace, we also use the same space-time signature $(\eta_{ab}) \equiv \text{diag.}\,(-,+,+)$.

$$F_{\alpha b}{}^I = -(\gamma_b\lambda^I)_\alpha \ , \quad F_{\alpha\beta}{}^I = 0 \ , \tag{5.4b}$$

$$\nabla_\alpha \lambda^{\beta I} = -\tfrac{1}{2}(\gamma^{cd})_\alpha{}^\beta F_{cd}{}^I \ , \tag{5.4c}$$

$$\nabla_\alpha e^\varphi = -\chi_\alpha e^\varphi \ , \tag{5.4d}$$

$$\nabla_\alpha \chi_\beta{}^I = +(\gamma^c)_{\alpha\beta}\Big[P_c{}^I - \tfrac{1}{4}f^{IJK}(\overline{\chi}^J\gamma_c\chi^K)\Big] \ , \tag{5.4e}$$

$$\nabla_\alpha P_b{}^I = -\nabla_b\chi_\alpha{}^I - [\chi_\alpha, P_b]^I - m(\gamma_b\lambda^I)_\alpha \ , \tag{5.4f}$$

$$\nabla_{[a}P_{b]}{}^I = +mF_{ab}{}^I + [P_a, P_b]^I \ , \tag{5.4g}$$

We can see that the transformation rules (3.5c) - (3.5f) can be recovered by the usual technique in superspace [18] relating to their component fields.

As usual, we can get also the component field equations with supercovariantized field strengths from the BIds at dimensions $d \geq 3/2$. These are superfield equations whose $\theta = 0$ sector correspond to component field equations, summarized as

$$(\slashed{\nabla}\lambda^I + m\chi^I)_\alpha \doteq 0 \ , \tag{5.5a}$$

$$(\slashed{\mathcal{D}}\chi^I)_\alpha + m\lambda_\alpha{}^I - \tfrac{1}{12}h^{IJKL}\chi_\alpha{}^K(\overline{\chi}^J\chi^L) \doteq 0 \ , \tag{5.5b}$$

$$\nabla_b F^{abI} - \tfrac{1}{2}mf^{IJK}(\overline{\lambda}^J\gamma^a\lambda^K) + mP^{aI} - \tfrac{1}{4}mf^{IJK}(\overline{\chi}^J\gamma^a\chi^K) \doteq 0 \ , \tag{5.5c}$$

$$\nabla_a P^{aI} + \tfrac{1}{2}f^{IJK}(\overline{\chi}^J\slashed{\mathcal{D}}\chi^K) - \tfrac{1}{8}h^{IJ,KL}(\overline{\chi}^K\gamma_a\chi^L)P_a{}^J \doteq 0 \ . \tag{5.5d}$$

We stress not only that these superfield equations obtained from BIds at $d \geq 3/2$ are consistent with our component results, but also that the mutual consistency among equations in (5.5) can be confirmed. For example, we can confirm that the $\nabla_a$-divergence of (5.5c) actually vanishes by the use of other superfield equations:

$$\nabla_a\Big[\nabla_b F^{abI} - \tfrac{1}{2}mf^{IJK}(\overline{\lambda}^J\gamma^a\lambda^K) + mP^{aI} - \tfrac{1}{4}mf^{IJK}(\overline{\chi}^J\gamma^a\chi^K)\Big]$$
$$= +mf^{IJK}(\overline{\lambda}^J\mathcal{F}_{(\lambda)}^K) - m\mathcal{F}_{(P)}^I + mf^{IJK}(\overline{\chi}^J\mathcal{F}_{(\chi)}^K) - \tfrac{1}{12}m^2 k^{IJ,L,MN}(\overline{\chi}^L\chi^M)(\overline{\chi}^L\chi^N) \ , \tag{5.6}$$

where

$$k^{IJ,K,LM} \equiv f^{IJN}h^{NK,LM} = f^{IJN}f^{NKP}f^{PLM} \ , \tag{5.7}$$

The $\mathcal{F}$'s corresponds respectively to the LHSs of (5.5a) through (5.5d), defined by

$$\mathcal{F}_{(\lambda)\alpha}{}^I \equiv -(\slashed{\nabla}\lambda^I + m\chi^I)_\alpha \ , \tag{5.8a}$$

$$\mathcal{F}_{(\chi)\alpha}{}^I \equiv -\Big[\slashed{\mathcal{D}}\chi^I + m\lambda^I - \tfrac{1}{12}h^{IJ,KL}\chi^K(\overline{\chi}^J\chi^L)\Big]_\alpha \ , \tag{5.8b}$$

$$\mathcal{F}_{(A)a}{}^I \equiv -\nabla_b F_a{}^{bI} + \tfrac{1}{2} m f^{IJK} (\overline{\lambda}^J \gamma_a \lambda^K) - m P_a{}^I + \tfrac{1}{4} m f^{IJK} (\overline{\chi}^J \gamma_a \chi^K) ~~, \quad (5.8c)$$

$$\mathcal{F}_{(P)}{}^I \equiv +\nabla_a P^{aI} + \tfrac{1}{2} f^{IJK} (\overline{\chi}^J \slashed{\mathcal{D}} \chi^K) - \tfrac{1}{8} h^{IJ,KL} (\overline{\chi}^K \gamma^a \chi^L) P_a{}^J ~~, \quad (5.8d)$$

Note that the last term in (5.6) vanishes identically:

$$k^{IJ,L,MN} (\overline{\chi}^J \chi^M)(\overline{\chi}^L \chi^N) \equiv 0 ~~, \quad (5.9)$$

as can be confirmed by Fierzing. Note that eq. (5.6) itself is rigorously an *identity* without the use of superfield equations. However, the superfield equations $(5.8a) \doteq (5.8b) \doteq (5.8d) \doteq 0$ combined with (5.9) yield the consistent result $(5.6) \doteq 0$.

There are other confirmations of the mutual consistency among the superfield equations, such as

$$\nabla_\alpha \mathcal{F}_{(\lambda)\beta}{}^I = +(\gamma^c)_{\alpha\beta} \mathcal{F}_{(A)c}{}^I \doteq 0 ~~, \quad (5.10)$$

where the $\mathcal{F}_{(A)c}{}^I \doteq 0$ (5.5c) is used only for the last equality. Similarly, we can confirm that

$$\nabla_\alpha \mathcal{F}_{(A)b}{}^I = +(\gamma_b \slashed{\nabla} \mathcal{F}_{(\lambda)}{}^I)_\alpha + m(\gamma_b \mathcal{F}_{(\chi)}{}^I)_\alpha - \nabla_b \mathcal{F}_{(\lambda)\alpha}{}^I \doteq 0 ~~, \quad (5.11)$$

where we have used $\mathcal{F}_{(\lambda)}{}^I \doteq 0$ and $\mathcal{F}_{(\chi)}{}^I \doteq 0$ only for the last equality.

As we have seen, in the BId formulation, we can get non-trivial relationships that are technically useful in superspace. These confirmations provide more than enough supporting evidence of the total consistency of our system, in particular, the non-trivial interplay between the vector multiplet and compensator scalar multiplet.

## 6. Concluding Remarks

In this paper we have completed the coupling of $N = 1$ supersymmetric self-dual Yang-Mills multiplet in 3D [2] to supergravity, including all the quartic terms. Thanks to the compensator formulation with manifest gauge symmetry, the coupling to supergravity is straightforward, like other supergravity formulations [4] for regular field strengths with Noether couplings. Before our previous paper [2] based on compensator formulation, such couplings had been thought to be extremely difficult, if not impossible [1]. We have given the general couplings between vector, compensator scalar and supergravity multiplets in (3.1), including the kinetic terms, topological mass terms with the supersymmetric Chern-Simons terms of the vector multiplet, together with the supersymmetric cosmological constant term. We stress that the completion of supergravity couplings including all the quartic fermionic terms provides the strong supporting evidence of the total consistency of our system at the classical level.

There can be three mass parameters possible in our system, namely, the non-Abelian coupling constant $m$, the Chern-Simons mass parameter $\mu$ quantized for groups with $\pi_3(H_0) \neq 0$, and the gravitino mass $M$ related to the cosmological constant $\Lambda = 2M^2$. For a non-Abelian gauge group $H_0$ with non-trivial $\pi_3(H_0)$, the parameter $\mu$ should be quantized as in (3.8). If we require one more mass relation, such as one of the masses of $\lambda$ or $\chi$ to be zero in certain bases, then the cosmological constant $\Lambda$ will be also quantized.

We have further performed the coupling of the self-dual Yang-Mills gauge field to the $\sigma$-model on the coset $G/H \equiv SO(8,n)/SO(8) \times SO(n)$. Namely, we have gauged an arbitrary subgroup $H_0 \subset H$ by the self-dual Yang-Mills vector with the arbitrary coupling constant $m$. To our knowledge, this is the first simple system that entertains the coupling of a self-dual gauge field in 3D to a gauged $\sigma$-model on the non-trivial coset $SO(8,n)/SO(8) \times SO(n)$ with arbitrary gauging for $^\forall H_0 \in H \equiv SO(8) \times SO(n)$.

We have also reformulated some of our component results in superspace. There seems to be no obstruction against coupling the self-dual supersymmetric Yang-Mills multiplet, based on compensator scalar multiplet. The compensator multiplet has provided a good framework to make these couplings straightforward. We have seen that the BIds at $d \geq 3/2$ provide all the superfield equations, and their mutual consistency can be confirmed by highly non-trivial relationships, including Fierzing of fermions such as (5.9). This superspace result also provides the supporting evidence of the classical consistency of our system.

Even though the number of supersymmetry is limited to $N=1$, nevertheless rich and non-trivial structures are seen to emerge in our model. This is due to general non-Abelian gauge groups we are dealing with, together with such topological properties as self-duality in 3D. We can expect more results for further generalizations of self-dual Yang-Mills fields in 3D to extended global or local supersymmetric models.